\documentclass[preprint,12pt]{elsarticle}

\usepackage{amssymb}
\usepackage{enumitem}

\usepackage{multirow}

\usepackage{subcaption}

\usepackage[hidelinks]{hyperref}

\newcommand{\full}{\ $\bullet$}
\newcommand{\prt}{\ $\circ$}

\begin{document}

\begin{frontmatter}



\title{ROSTAM: A Passwordless Web Single Sign-on Solution Mitigating Server Breaches and Integrating Credential Manager and Federated Identity Systems}

\author[informatics]{Amin Mahnamfar}
\ead{a.mahnamfar@gmail.com}

% \affiliation[label2]{organization={},
\author[informatics,computer,securify]{Kemal Bicakci}
\ead{kemalbicakci@itu.edu.tr}
\author[securify]{Yusuf Uzunay}
\ead{yusuf.uzunay@securifyidentity.com}
\affiliation[informatics]{organization={Informatics Institute, Istanbul Technical University},
           city={Istanbul},
           postcode={34467},
           country={Türkiye}}
\affiliation[computer]{organization={Computer Engineering Department, Istanbul Technical University},
           city={Istanbul},
           postcode={34467},
           country={Türkiye}}
\affiliation[securify]{organization={Securify Information Technology and Security Training Consulting Inc.},
            city={Ankara},
            postcode={06378},
            country={Türkiye}}

\begin{abstract}
The challenge of achieving passwordless user authentication is real given the prevalence of web applications that keep asking passwords. Complicating this issue further, in an enterprise environment,  a single sign-on (SSO) service is often maintained but not all applications can be integrated with it. We envision a passwordless future which provides a frictionless and trustworthy online experience for users by integrating credential management and federated identity systems. In this regard, our implementation ROSTAM offers a dashboard that presents all applications the user can access with a single click after a passwordless SSO. The security of web passwords on the credential manager is ensured with a Master Key, rather than a Master Password, so that encrypted passwords can remain secure even if stolen from the server. We propose and implement novel techniques for synchronization (pairing) and recovery of this Master Key. We compare our solution to previous work using different evaluation frameworks, demonstrating that our hybrid solution combines the benefits of credential management and federated identity systems.

\end{abstract}



\begin{keyword}


single sign-On \sep SSO \sep password managers \sep usable security \sep  master password \sep master key \sep passwordless \sep user authentication
\end{keyword}

\end{frontmatter}

% \linenumbers

\section*{Abbreviations}

\begin{tabular}{ll}
{\bf{2FA}} & {\bf:} Two-Factor Authentication\\
{\bf{AES}} & {\bf:} Advanced Encryption Standard\\
{\bf{API}} & {\bf:} Application Programming Interface\\
{\bf{CM}} & {\bf:} Credential Manager\\
{\bf{CTR}} & {\bf:} Counter Mode (encryption)\\
{\bf{FIS}} & {\bf:} Federated Identity Systems\\
{\bf{ID}} & {\bf:} Identification Number\\
{\bf{IdP}} & {\bf:} Identity Provider\\
{\bf{MAC}} & {\bf:} Message Authentication Code\\
{\bf{OAEP}} & {\bf:} Optimal Asymmetric Encryption Padding\\
{\bf{PRNG}} & {\bf:} Pseudorandom Number Generator\\
{\bf{PrivKey}} & {\bf:} Private Key\\
{\bf{PubKey}} & {\bf:} Public Key\\
{\bf{PubKeyHash}} & {\bf:} A hashed version of a Public Key\\
{\bf{QR Code}} & {\bf:} Quick Response Code\\
{\bf{RP}} & {\bf:} Relying Party\\
{\bf{RSA}} & {\bf:} Rivest–Shamir–Adleman\\
{\bf{SHA}} & {\bf:} Secure Hash Algorithm\\
{\bf{SP}} & {\bf:} Service Provider\\
{\bf{SSO}} & {\bf:} Single Sign-On\\
{\bf{TPM}} & {\bf:} Trusted Platform Module\\
{\bf{URL}} & {\bf:} Uniform Resource Locator\\
\end{tabular}

\section{Introduction}
\label{sec:intro}
The endeavor to eliminate passwords appears to be an ongoing and protracted journey. Considering the deployability and cost benefits attributed to passwords for websites, an abrupt replacement with an alternative scheme appears impracticable. Nevertheless, the exploration of novel solutions aimed at ameliorating the substantial limitations inherent in passwords remains viable. A comparison of web authentication alternatives showed that credential managers (CM) and federated identity systems (FIS) are currently two of the most promising solutions in this regard as compared to other schemes \cite{Replacepasswords}. In a more recent work, Alaca and Van Oorschot had a closer look to the available CM and FIS~\footnote{Both were considered as single sign on systems (SSO) by Alaca and Van Oorschot. On the other hand, the term SSO is usually reserved for FIS use cases by others. In our paper, we will try to avoid ambiguity regarding these two different uses as much as possible.} options and pointed out the benefits of each \cite{SSOAnalysis}. 
%Of course, it is not possible to choose one scheme over another but it is possible to come up with a hybrid scheme which improves one another’s drawbacks. We will discuss the details in the next part of the article.

In our work, primarily, an enterprise setting is considered where we have already embraced a FIS solution. On the other hand, as expected, not every single service provider (web application) could be integrated to it. There is certainly a need for a CM if we do not want our employees to manage their passwords by themselves (perhaps due to security reasons). Instead of having independent FIS and CM systems in operation, we design and implement a CM (basically as a Browser Extension alongside a mobile app) which gets a service from the already available identity provider(IdP) using OpenID Connect protocol. 

In our hybrid solution, users login to the SSO system using their mobile app without a need to enter a password and could start using both CM and FIS parts seamlessly. Our goal in the design of the CM part is to make it resistant against server compromise without using a master password users should manage. In our system called ROSTAM, we put forward an innovative solution where credentials are secured with a Master Key instead of a Master Password. The Master Key is stored not on the server but on user's mobile phone. In case the mobile phone is not available, the Master Key is recoverable if users have a reach to at least one of the browsers with an installed ROSTAM extension. Up to our best knowledge, this hybrid design of ROSTAM integrating FIS and CM approaches together with bringing both security (passwords are not recoverable even when servers are compromised) and usability (no need to memorize even a single password) benefits is novel and represents a pioneering approach in the field.

The rest of the paper is organized as follows. In section \ref{sec:related}, we review the previous work. In section \ref{sec:design}, we present the design and implementation of our solution ROSTAM. In section \ref{sec:security}, we make an informal security analysis of our system and consider possible internal and external threats and how secure ROSTAM is against them. In section \ref{sec:comparison}, we take a look at existing proposals such as using a Master Password and compare it to our scheme, pointing out the strength and extra benefits ROSTAM provides. In section \ref{sec:evaluation}, we compare  ROSTAM with other prominent solutions in this space and examine what improvements our hybrid SSO system provides. We present the limitations of ROSTAM together with some promising future work directions in section \ref{sec:limitations} and we conclude our paper in section \ref{sec:conclusion}.

\section{Related Work}
\label{sec:related}

Pashalidis et al. \cite{SSOCategories} introduced a taxonomy for existing Single Sign-On (SSO) systems, categorizing them into two main groups: pseudo-SSO and true-SSO systems. They also divided each category to Local and Proxy-based systems. In a more recent work, Alaca et al. \cite{SSOAnalysis} developed a framework and identified 14 benefits in security, usability, deployability and privacy categories in order to evaluate 12 web SSO systems deployed and/or proposed. We also compare our proposal to existing work using this framework (in section 6). 
%In our system we try to improve the lacking benefits of a widely adopted schemes with also maintaining the existing benefits of them. We follow the best schemes from the category of credential managers (Firefox Sync 2.0) and federated identity systems (OpenID Connect) based on the evaluation table done in the previous work in order to develop a hybrid scheme having both systems’ benefits.

Alternatives to password based user authentication is an extensively studied topic. We review only a selected subset of these alternatives here, for a more comprehensive treatment of the subject, we refer readers to seminal work by Bonneau et al. \cite{Replacepasswords}. 

%Based on this our solution does not fall specifically under any of these four categories as our SSO supports both legacy-based authentication (e.g., Firefox Sync 2.0) and authentication assertions (e.g., OpenID Connect). In addition, we cannot consider Rostam completely Local or Proxy-based since user is required to authenticate himself to a remote server and on the hand user’s data is kept in his side locally. Rostam also compensates disadvantages which exist in these categories individually.

The quest to replace password is still continuing and CMs offer improvements while supporting legacy-based servers, without requiring any major changes, but majority of them provides protection with a user-chosen master password and this does not totally eliminate security risks of passwords. One notable exception is the work by McCarney et al. \cite{tapas}, which proposed a password manager protecting passwords with a strong encryption key. Smartphones become inseparable part of life. With that in mind, the proposed password manager called Tapas uses smartphones to offer a dual-possession authentication system in a way that for an attacker to successfully break the system he is required to have access to more than one device. Tapas also uses an out-of-band method for pairing process using QR codes.
Although being a promising candidate to eliminate passwords, one of the main deficiencies of Tapas is the lack of password recovery in case the mobile phone is not available. Besides, Tapas does not offer any capability regarding federation identity systems.

%\begin{itemize}
%\item Tapas uses online server just for establishing P2P connection but we use it for querying users’ encrypted data whenever required by cellphone or the extension
%\item Tapas keeps the encrypted data on the cellphone but we keep it in our cloud servers in an encrypted from.
%\item Tapas requires user to browse the website manually and then open his/her cellphone, tap on the related website and account which then extension gets it and decrypts the data and logs user in.
%\item Our scheme includes a recovery process which Tapas lacks.
%\item Tapas only works with just one cellphone and browser extension but in Rostam users can pair and add as many as devices they want.
%\end{itemize}

In 2010, Karole et al. \cite{OnlinevsofflineCMUserstudy} conducted a user study regarding choices between portable and online password managers. Surprisingly, most users preferred the former especially cellphone-based ones due to trust related reasons despite the fact that online password managers could be more usable. In our work, we design CM part of ROSTAM so that servers only store encrypted user passwords and do not have access to the keys. An attacker needs to breach both the server and one of the paired devices to compromise passwords.

\section{Design and Implementation}
\label{sec:design}
ROSTAM is hybrid SSO system using a browser extension companion which works as an online Credential Manager(CM). The extension is basically just one of the service providers(SPs) which relies on the authentication service of IdP to grant access to users. 

\begin{figure}
\includegraphics[width=\columnwidth]{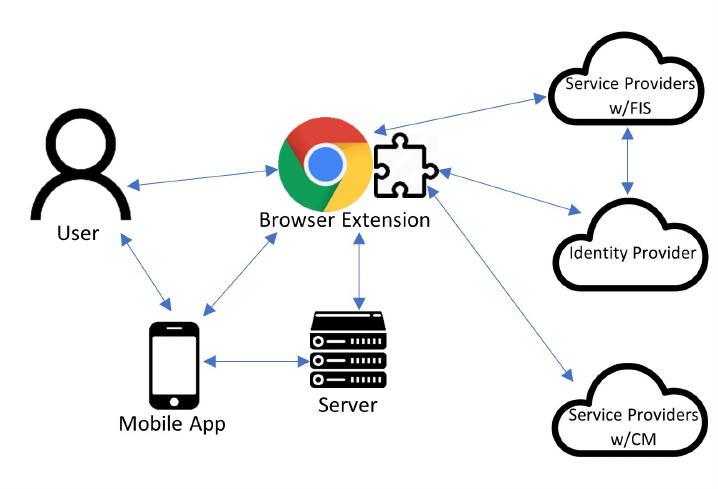}
\centering
\small
\caption{System model of ROSTAM.}
\label{fig:components}
\end{figure}

In our system model, there are six main elements. Below, we provide a brief explanation for each of them:

\begin{itemize}
    \item \textbf{User:} User is an individual (employee) who uses ROSTAM to access his/her services or resources. Users can create and manage their accounts, reset passwords, and authenticate themselves to gain access to protected resources.
    \item \textbf{Service Providers:} Service Providers(SPs) are organizations or applications that provide services to users. These services can range from online shopping to social media or enterprise software. These SPs are divided into two categories, the ones using FIS which are also called Relying parties (RPs) and the ones users authenticate with either manual password entry or using ROSTAM's CM.
    \item \textbf{Browser Extension:} A software component (developed for Chrome) that adds the functionality regarding all user tasks in ROSTAM to the web browser.
    \item \textbf{Identity Provider:} A trusted party that authenticates users by SSO (implemented with the OpenID Connect Protocol) and provides authentication and other services to RPs.
    \item \textbf{Server:} It is the backend system that hosts the identity management system and stores user data, application configurations, and other system information. The server is also responsible for tasks such as processing user requests, validating user credentials, and enforcing security policies.
    \item \textbf{Mobile App:} An application installed on a mobile device (running Android 7 or higher) and used for tasks such as generation of the encryption key and pairing with the Extension using QR codes. It also acts as a mobile password manager.
\end{itemize}
Figure \ref{fig:components} shows the system components and how they interact. More details will be provided in the following subsections.

Before explaining the design details, we would like to give information about our implementation choices. The Extension is developed mostly in pure JavaScript and for parts such as user interface and server requests, jQuery methods \cite{JQuery} are used. For the styling and user interface, Bootstrap framework \cite{Bootstrap} is used. 

OpenID Connect protocol is used in order to authenticate users to the IdP which gives service both to the Extension and to the other registered SPs. For the server-side or backend part, Python Flask web framework  \cite{Flask} is used. Because of this choice, a suitable library for the OpenID Connect \cite{OpenID} implementation is Authlib \cite{Authlib}, a certified OAuth 2.0 \cite{OAuth} library. The Extension utilizes two databases: a local one called IndexedDB, which resides in Google Chrome \cite{Chrome} and implemented using a library called Dexie.js \cite{Dexie} and a remote one, MongoDB \cite{MongoDB}, which is used for backup and synchronization of the passwords through different devices.  

In order to be able to use ROSTAM, users are currently required to use an Android \cite{Android} phone for the Mobile App and also use Google Chrome (or other Chromium-based) web browser with the Extension installed on any supported desktop operating system. Users are also required to carry their cellphone having a persistent Internet connection.

The rest of this section is divided into six subsections. We first explain initial setup phase. Then, we explain how login works. In the third subsection, the pairing with a new device is explained. Then, we explain how ROSTAM can be used as an IdP supporting OpenID Connect protocol for SPs. In the fifth subsection we focus on the CM properties of ROSTAM. Finally, we take a look at the inner details of our innovative solution for account recovery.

\subsection{Initial Setup}
\label{sec:setup}

We divide the required initial setup process to three different steps:
\begin{enumerate}[label=\arabic*., start=1]
  \item Mobile App setup
  \item Mobile authentication and association with ROSTAM's IdP
  \item Browser Extension setup
\end{enumerate}
The first two steps only happen once (as long as User owns the same mobile device). The last step gets repeated for any new browser the User wants to use.
\begin{enumerate}[label=\arabic*., start=1]
  \item Mobile App setup:
  \label{subsec:mobilesetup}
 Since ROSTAM is designed mainly for an enterprise setting, we assume that the company adapting ROSTAM has already established a centralized directory such as Microsoft Active Directory for its employees and the User is already registered as one of these employees. This means that user information such as email address and cellphone number is available. So, after downloading and installing the Mobile App from the respective store e.g., PlayStore \cite{PlayStore}, an employee can register himself/herself as ROSTAM User. After the initial authentication \footnote{Depending on security policy of the enterprise, this initial authentication could be implemented in various ways such as email confirmation (not secure), face recognition or in-person verification.} leading to a successful registration, a random encryption key which we refer to as MasterKey and also a random key pair which is referred as PubKey and PrivKey gets created on the Mobile App. The later ones are RSA keys which get created and stored directly in local KeyStore \cite{AndroidKeyStore}. They are used for Encryption and Decryption of data transferred between the Mobile App and the remote database on the Server. MasterKey is generated with a procedure described as follows:
  \begin{enumerate}[label=\arabic{enumi}.\arabic*., start=1]
    \item A 256-bit random AES key gets generated using Java crypto library \cite{JavaCrypto} outside the KeyStore.
    \item The generated key gets encrypted with the PubKey of the cellphone.
    \item The Mobile App calculates hash (SHA-256 \cite{SHA}) of PubKey and transfers PubKey, PubKey hash and encrypted MasterKey to the remote database.
    \item Finally, Mobile App imports MasterKey to the KeyStore making it unexportable.
  \end{enumerate}
  \item Mobile authentication and association with ROSTAM's IdP:
  \label{subsec:idpsetup}
  We use passwordless authentication for this purpose (the initial authentication during the Mobile App setup and registration expires after a pre-determined duration and authentication should be repeated):
  \begin{enumerate}[label=\arabic{enumi}.\arabic*., start=1]
    \item The User enters username / email address on the phone.
    \item ROSTAM sends a request to the company's Active Directory server to retrieve the User's public key associated with the entered email address.
    \item The User receives a push notification on the cellphone.
    \item The User confirms the login attempt by providing biometrics (e.g., fingerprint).
    \item The Mobile App generates a signed assertion and sends it back to ROSTAM's Server.
    \item Rostam verifies the signed assertion with the User's public key.
    \item If the verification is successful, authentication is achieved and ROSTAM grants the Mobile App access to the system.
  \end{enumerate}
  \item Browser Extension setup
  \label{subsec:extensionsetup}
  Before using the Extension, User is required to install it from Chrome Store \cite{ChromeWebStore}. No extra step is required.
\end{enumerate}

\subsection{Authentication and log-in}
\label{sec:authentication}
To be able to use ROSTAM to login to any web app, the User needs to authenticate himself to ROSTAM (IdP) from the browser. This is similar to mobile authentication explained above but this time the User pays a visit to the login page of SSO either by manually entering the URL on the browser or by clicking the login button in the Extension.

%After successful authentication, by clicking on the extension, it checks in the background for an existing session with the help of OpenID Connect protocol and user automatically gets logged into the extension also. Here are the steps which shows how this flow works in detail (Figure~\ref{fig:authentication}):
%\begin{enumerate}
%    \item User pays a visit to the log in page of SSO by either manually entering the URL or clicking the log in button in the extension.
%    \item User is required to input his username (email or phone number) and then press log in button.
%    \item A push notification will be sent to user’s trusted mobile phone (user must have %authenticated himself already using the app).
%    \item After approving the session, in push notification, user will be logged in.
%\end{enumerate}

%\begin{figure}
%\includegraphics[width=\columnwidth]{res/Authentication.pdf}
%\centering
%\small
%\caption{Authentication steps between user and Rostam described in~\ref{sec:authentication}}
%\label{fig:authentication}
%\end{figure}

\subsection{Pairing}
\label{sec:pairing}

\begin{figure}
\includegraphics[width=\columnwidth]{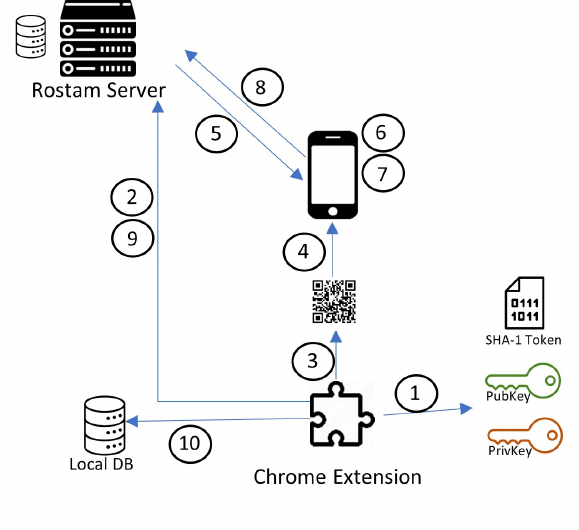}
\centering
\small
\caption{Pairing a new device in ROSTAM.}
\label{fig:pairing}
\end{figure}

\begin{figure}
    \includegraphics[width=8cm,keepaspectratio=true]{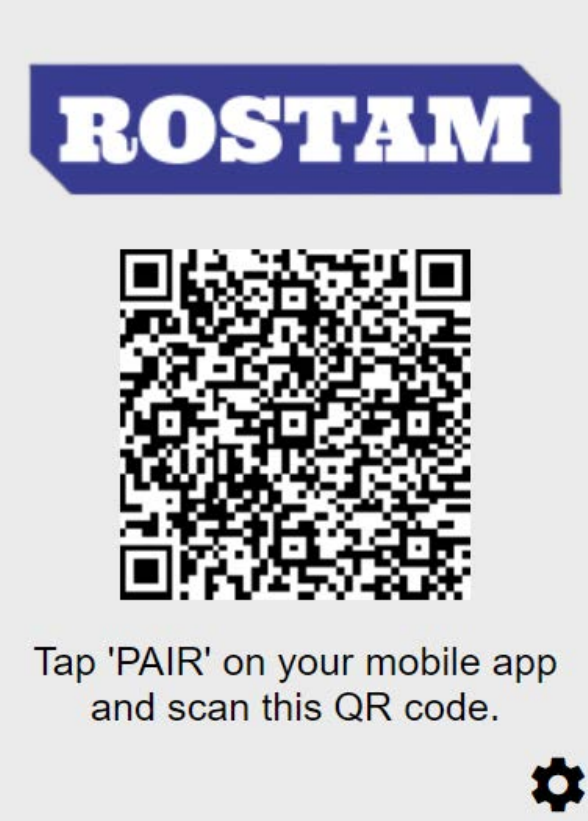}
    \centering
    \small
    \caption{QR Code generated from a token and PubKeyHash displayed by the Extension for pairing .}
    \label{fig:QRCodePair}
\end{figure}

After successful completion of the setup and authentication steps, the User is required to initiate also a pairing process for once in any new browser which Browser Extension gets installed. The purpose of pairing is to have a secure setup for the transfer of credentials to the device browser is running on. In this process we take advantage of using QR codes to put less burden on user while preserving privacy and security. In case the User has already been authenticated, the pairing process proceeds as follows (see Figure~\ref{fig:pairing}):
\begin{enumerate}
  \item A pair of keys called PubKey and PrivKey (RSA 2048-bit/OAEP) and also a random (PRNG algorithm) 128-bit one time token\footnote{Smartphone can verify PubKey of the Extension as it is scanned through camera but it is not feasible to do the other way around, so we use this token to verify the presence of the genuine smartphone. Refer to section~\ref{sec:ssthreats} for more details.} gets created and saved on the device in local DB making PrivKey unexportable.
  \item The Extension sends PubKey and its hash (SHA-256) to the Server in order to get it saved in remote database. It also checks continuously for a MasterKey field update in the relevant location of the remote database.
  \item QR code of token concatenated with the hashed version of a Public Key(PubKeyHash) is shown by the Extension (see Figure~\ref{fig:QRCodePair}).
  \item User clicks on the "Pair" button on the Mobile App and scans this code with the mobile phone.
  \item Mobile App splits the token and PubKey hash and queries the Extension’s PubKey and encrypted MasterKey from the Server.
  \item Mobile App verifies the received PubKey by calculating its SHA-256 hash and comparing it to the PubKey hash. If they are same then it decrypts the encrypted MasterKey with its PrivKey in KeyStore.
  \item Mobile App concatenates the token with the MasterKey encrypted with the newly paired device’s PubKey obtained earlier.
  \item Mobile App adds the encrypted token and the MasterKey to the corresponding location in the remote database.
  \item The Extension detects the added data (continuous check has already started in step 3) and gets it.
  \item The Extension performs the decryption with its PrivKey. It splits the token from MasterKey and then compares it with the one in local database. If they match, then MasterKey gets imported into the IndexedDB database making it unexportable.
\end{enumerate}

After these steps, our system becomes fully functional and ready for use.

\subsection{SPs supporting OpenID Connect}
\label{sec:pbauthentication}
SPs, also called RPs, which support OpenID Connect protocol, are required to register with ROSTAM and accept it as the IdP in order to make it possible for users to login with SSO. This SSO functionality (in its traditional meaning) mainly targets enterprise applications. SPs can be added by the organization or by users themselves.

\begin{figure}
\includegraphics[width=\columnwidth]{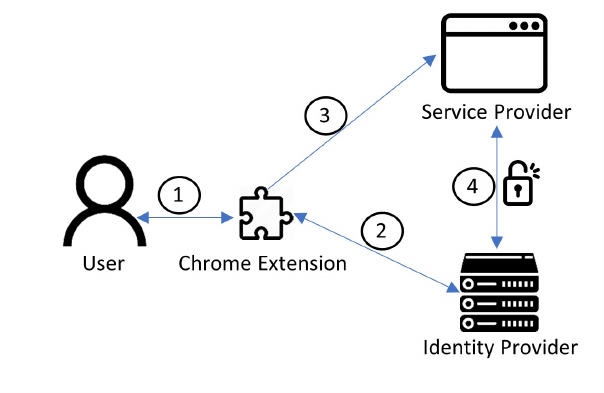}
\centering
\small
\caption{User authentication for SPs supporting OpenID Connect protocol.}
\label{fig:pbauthentication}
\end{figure}

To exemplify, consider \textit{example.com} as the ROSTAM’s domain name, the actions required to be taken so that users could authenticate themselves to a SP with ROSTAM's IdP are as follows:
\begin{enumerate}
    \item Redirect and logout URLs are configured.
    \item A Client ID and Client Secret gets generated. The SP is required to use these values while communicating with ROSTAM over OpenID Connect protocol.
    \item The other required endpoints for OpenID Connect protocol are defined as in \textit{https://example.com/.well-known/openid-configuration} so the SP can fetch them all with setting \textit{server\_metadata\_url} to the above URL.
\end{enumerate}
After these configurations have been made, the User can log in to the RPs as follows (Figure~\ref{fig:pbauthentication}):
\begin{enumerate}
  \item User clicks on the Extension.
  \item The Extension checks whether a live authenticated session already exists or not by communicating with ROSTAM's Server. If there is a live session then it redirects the User to a page which shows icons for already registered user accounts. If not, it redirects User to the login page.
  \item When User clicks on the icon of a RP, the desired web application opens in a new tab. 
  \item In the background, the relying party communicates with the IdP which provides the assertion about User's identity using OpenID Connect protocol. After the assertion is verified, the User gets authenticated to the desired RP.
\end{enumerate}

\subsection{Authentication to SPs with CM}
\label{sec:fbauthentication}
For an well-organized presentation of ROSTAM’s CM, we get help from the design paradigms by Simmons et al. \cite{PMUseCases}. Simmons et al. pointed out seven essential, three recommended and four extended use cases for password managers. For simplicity, we took only essential use cases into consideration. We note that in some use cases implementation of all paradigms at the same time is impossible due to their nature. Table \ref{table:design} shows the implemented use cases. We provide the details for these use cases as follows:

%Yet another group of paradigms are not currently implemented but we plan to implement and make them available in the next version of ROSTAM.

% Further explanation of why these paradigms are not suitable for our design is provided in the Appendix.

\begin{table*}[]
\centering
\resizebox{\textwidth}{!}{%
\begin{tabular}{ll}
\cline{1-2}
 \hline
 \textbf{Use Cases} & \textbf{Paradigm} \\
 \hline
 % \hyperref[subsec:mobilesetup]{(E1-P1) Install an app}
 \multirow{3}{*}{(E1) Setup manager} & (E1-P1) Install an app\\
 & (E1-P2) Install an extension \\
 & (E1-P3) Requires a cloud account \\
 \hline
\multirow{2}{*}{(E2) Register credential} & (E2-P1) Manual registration \\
 & (E2-P2) Auto-detect registration \\
 \hline
 \multirow{2}{*}{(E3) Update credential} & (E3-P1) Manual update \\
 & (E3-P2) Auto-detect update \\
 \hline
 \multirow{1}{*}{(E4) Remove credential} & (E4-P1) Manual removal \\
 \hline
 \multirow{4}{*}{(E5) Autofill credential} & (E5-P1) Autofill w/ interaction \\
 & (E5-P2) Autofill w/o interaction \\
 & (E5-P3) Internal login tool \\
 & (E5-P4) Separate subdomains \\
 \hline
 \multirow{2}{*}{(E6) Manually enter credential} & (E6-P1) Show stored credentials \\
 & (E6-P2) Obfuscate password characters \\
 \hline
 \multirow{1}{*}{(E7) Generate password} & (E7-P1) Manual generation \\
 \hline
 \multirow{1}{*}{(E8) Sync credentials} & (E8-P1) Fully automated sync \\
 \hline
 \multirow{2}{*}{(E9) Lock manager} & (E9-P1) Manual lock \\
 & (E9-P2) Timed auto-lock \\
 \hline
 \multirow{2}{*}{(E10) Unlock manager} & (E10-P1) Unlock with master password \\
 & (E10-P2) Unlock with 2FA \\
 \hline

\end{tabular}%
}
\caption{Essential Use Cases and Paradigms (introduced by Simmons et al. \cite{PMUseCases}) implemented in ROSTAM.}
\label{table:design}
\end{table*}

\subsubsection{Setup Manager}
We have already discussed this use case, but in order to have a full conformance to Simmons et al.'s taxonomy \cite{PMUseCases}, we repeat it here as well. As we mentioned in \ref{sec:setup} , the User is required to install our mobile application (E1-P1) and the Browser Extension (E1-P2). The account on the Server (Cloud) also gets automatically created just after the User registers the Mobile App and is in continuous use while using the Extension (E1-P3). To complete the setup, the User is also required to pair with the Extension which is explained in \ref{sec:pairing}. For more details, we refer back to the aforementioned section.

\subsubsection{Register Credentials}
\label{sec:registerCredentials}
\begin{figure}[b]
    \includegraphics[width=8cm,keepaspectratio=true]{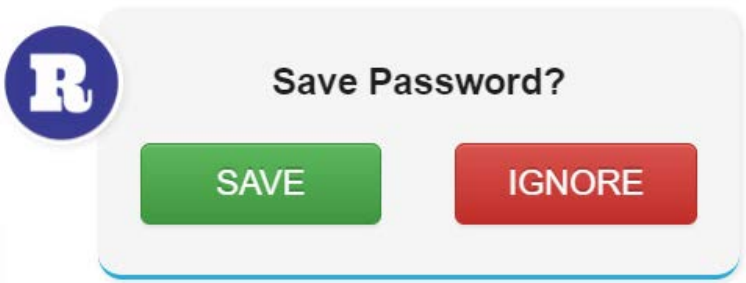}
    \centering
    \small
    \caption{Credential saving prompt.}
    \label{fig:confirmation}
\end{figure}

There are two ways a User can register his/her credentials. In the first method, when the User manually logins to a website, the Extension detects submitting new input and captures the credentials and then asks the User whether s/he wants to save it or not by showing a pop-up window (see Figure~\ref{fig:confirmation}). In case User selects “Yes”, the credentials gets saved (E2-P2).

\begin{figure}
    \includegraphics[width=\columnwidth,keepaspectratio=true]{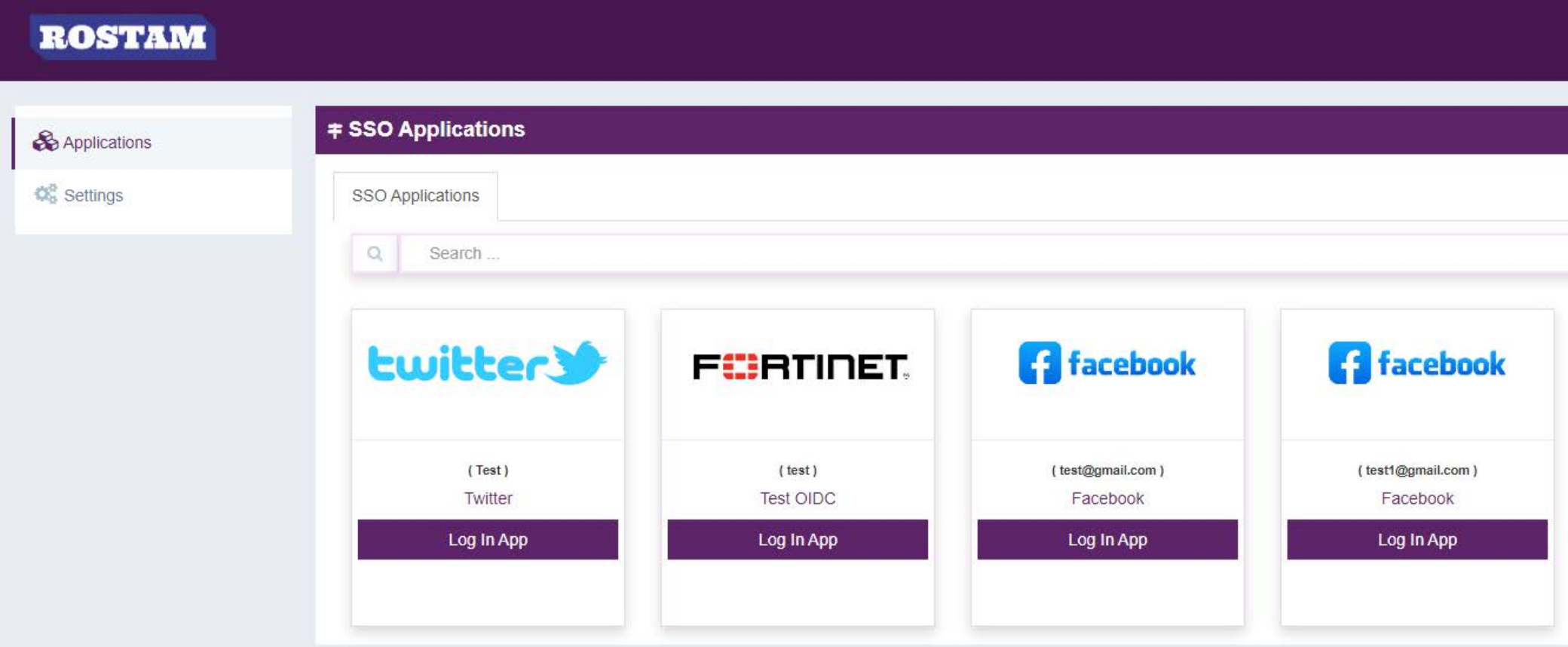}
    \centering
    \small
    \caption{SSO application panel after a User logs in.}
    \label{fig:SSOpanel}
\end{figure}

In the second method, User (or an admin) adds an application and its credentials manually to his/her account (see Figure~\ref{fig:SSOpanel}) (E2-P1). In addition to username and password, the User is also required to enter the SP’s login page URL. 

In both of these methods, the Extension uses the MasterKey to encrypt website’s login URL, username and password (with CTR mode and no padding) and append a Message Authentication Code(MAC). Then, it saves these in the Server.

%    \item Then, it saves the encrypted form of website, username and password to ROSTAM's server.

%The rest is same for both methods after clicking the Save button (Figure~\ref{fig:addSP}) described as follows:

%\begin{enumerate}
%    \item The extension uses the MasterKey to encrypt website’s login URL, username and password (with CTR mode and no padding) and separately append an HMAC.
%    \item Then, it saves the encrypted form of website, username and password to ROSTAM's server.
%\end{enumerate}

%\begin{figure}
%    \includegraphics[width=\columnwidth]{res/addSP.pdf}
%    \centering
%    \small
%    \caption{Storing passwords in ROSTAM.}
%    \label{fig:addSP}
%\end{figure}

\subsubsection{Update Credentials}
\label{sec:changePW}
Similar to registering, there are two ways for update. A User can manually update password in his account (E3-P1) or in case User visits login page of the SP and inputs a different password for the same username, he will get a prompt asking whether he wants to update the existing password. With the ''Yes” answer, the password gets updated (E3-P2). 

%The extension follows the steps below (Figure~\ref{fig:changePW}):
%\begin{enumerate}
%    \item After typing/capturing the new password, the user clicks on the Save button.
%    \item The extension captures and encrypts the new password with the MasterKey and also calculates and appends HMAC to it.
%    \item The extension sends the ID and encrypted password to the ROSTAM's server.
%    \item The server replaces the old encrypted password and HMAC with the new ones.
%\end{enumerate}
%\begin{figure}
%    \includegraphics[width=\columnwidth]{res/changePW.pdf}
%    \centering
%    \small
%    \caption{Changing passwords in ROSTAM.}
%    \label{fig:changePW}
%\end{figure}

\subsubsection{Remove Credentials}
\label{sec:removeCredential}
In the current version of ROSTAM, removal of a credential can be done only manually through the User's account (E4-P1).

\subsubsection{Autofill Credentials}
\label{sec:autofillCredentials}
User can login in two ways. He can either select the SP and the desired account from the Extension (see Figure~\ref{fig:extensionAccountList}) or visit the site directly (by typing the address in URL bar or by using a search engine). For a better understanding, the technical details are presented separately for these two cases as below:
\begin{figure}
    \includegraphics[width=8cm,keepaspectratio=true]{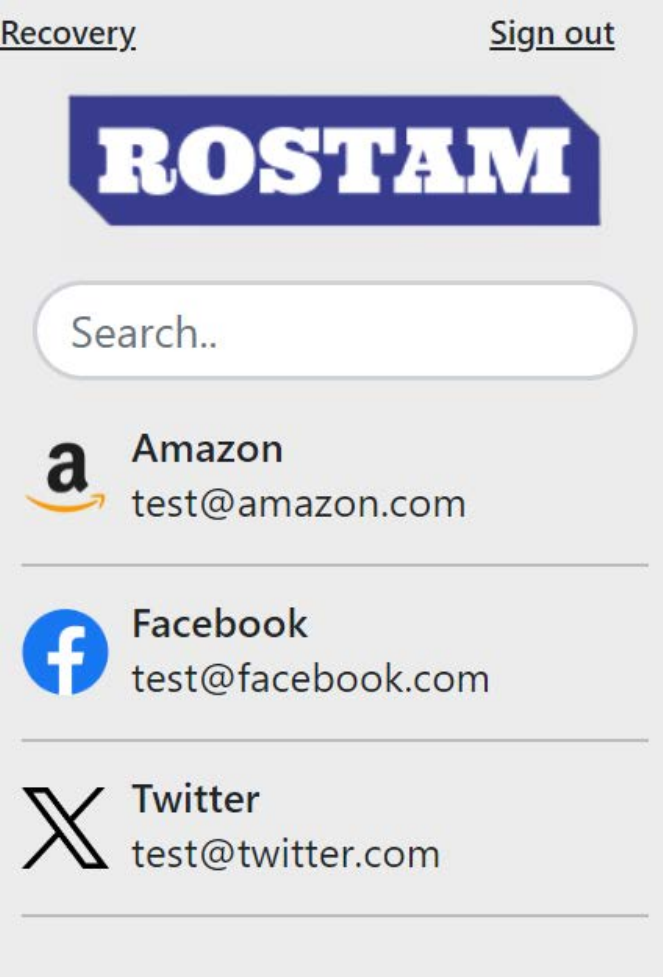}
    \centering
    \small
    \caption{User's account list displayed by the Extension.}
    \label{fig:extensionAccountList}
\end{figure}
\begin{enumerate}[label=\arabic*., start=1]
    \item When a User clicks on a SP from the Extension (E5-P3) with the desired username or account name from the shown list %(Figure~\ref{fig:fbauthentication1}):
    \begin{enumerate}[label=\arabic{enumi}.\arabic*., start=1]
        \item A new tab gets opened in the browser.
        \item The Extension sends a query with a unique ID  (generated when the SP gets stored for the first time in the remote database) to the Server.
        \item The Server returns the encrypted username and password corresponding to the ID.
        \item The Extension then decrypts the received data with the MasterKey and checks the MAC.
        \item It fills the required fields.
        \item The User clicks the login button (E5-P2).
    \end{enumerate}
    
%    \begin{figure}
%        \includegraphics[width=\columnwidth]{res/fbauthentication1.pdf}
%        \centering
%        \small
%        \caption{Autofilling credentials when extension is used.}
%        \label{fig:fbauthentication1}
%    \end{figure}
    \item When User visits a website manually by typing in the URL %(Figure~\ref{fig:fbauthentication2}):
    \begin{enumerate}[label=\arabic{enumi}.\arabic*., start=1]
        \item The Extension captures the website URL and encrypts it with the MasterKey.
        \item The Extension sends a query to the Server searching for usernames and passwords associated to that website. The URL needs to be an exact match, meaning that for a domain such as \textit{https://www.example.com/login}, the returned result should not include subdomains such as \textit{https://www.sub.example.com/login} (E5-P4).
        \item The Extension decrypts the username(s). If there is only one record, it goes to the next step, and if not, it shows the list of usernames in a pop-up window, asking the User to select one.
        \item The Extension decrypts the password for the (selected) username. It fills the required fields in the browser and the User is required to click on the login button (E5-P1)\footnote{Automatic clicking on the Login button is not preferred since it can be confusing for the User not expecting it. Manual click gives a last chance to the User to change things. For example, the User might have changed his/her password and might want to login with new credentials.}.
    \end{enumerate}
%    \begin{figure}
%        \includegraphics[width=\columnwidth]{res/fbauthentication2.pdf}
%        \centering
%        \small
%        \caption{Autofilling credentials when extension is not ,5used.}
 %       \label{fig:fbauthentication2}
 %   \end{figure}
\end{enumerate}

It is worth mentioning that data received from the Server is not stored in the local storage neither in encrypted form nor after decryption.

\subsubsection{Manually Enter Credentials}
\label{sec:manuallyEnterCredential}
\begin{figure}
    \centering

    \begin{subfigure}{0.32\textwidth}
        \centering
        \includegraphics[width=0.9\linewidth,keepaspectratio=true]{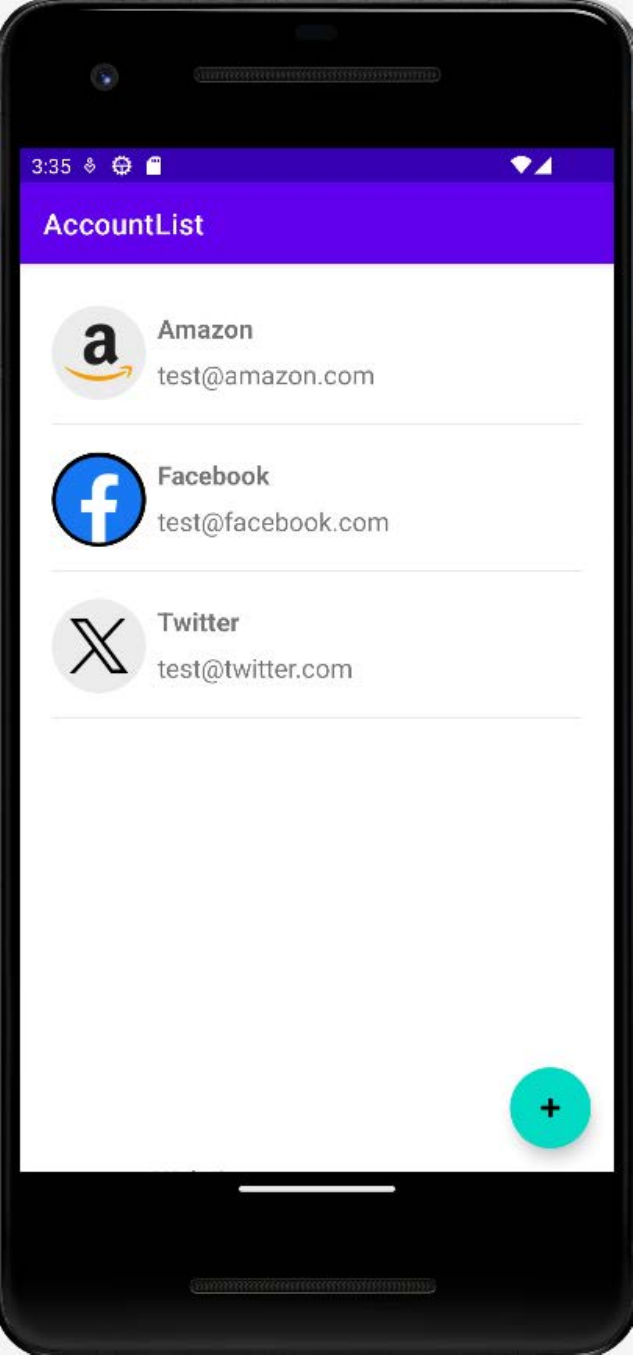}
        \caption{All saved SPs displayed on the mobile application of ROSTAM.}
        \label{fig:mobileAccountList}
    \end{subfigure}
    \hfill
    \begin{subfigure}{0.32\textwidth}
        \centering
        \includegraphics[width=0.9\linewidth,keepaspectratio=true]{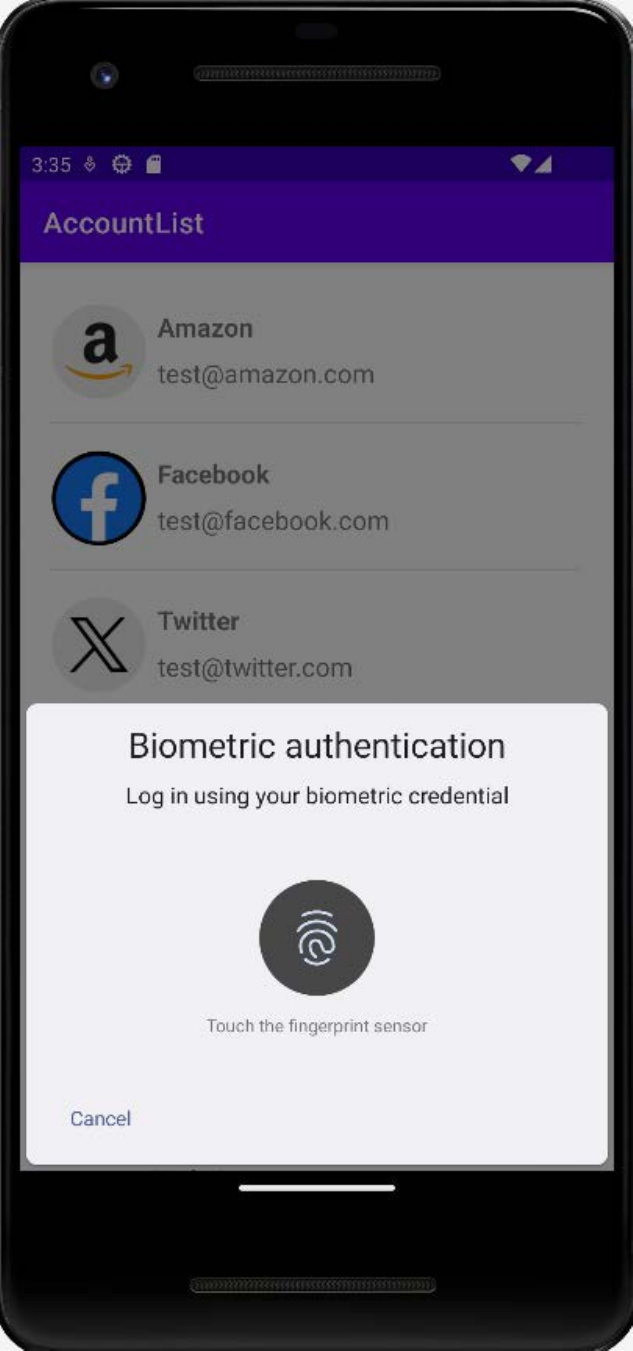}
        \caption{Biometric authentication request before showing the password for selected SP.}
        \label{fig:biometricApproval}
    \end{subfigure}
    \hfill
    \begin{subfigure}{0.32\textwidth}
        \centering
        \includegraphics[width=0.9\linewidth,keepaspectratio=true]{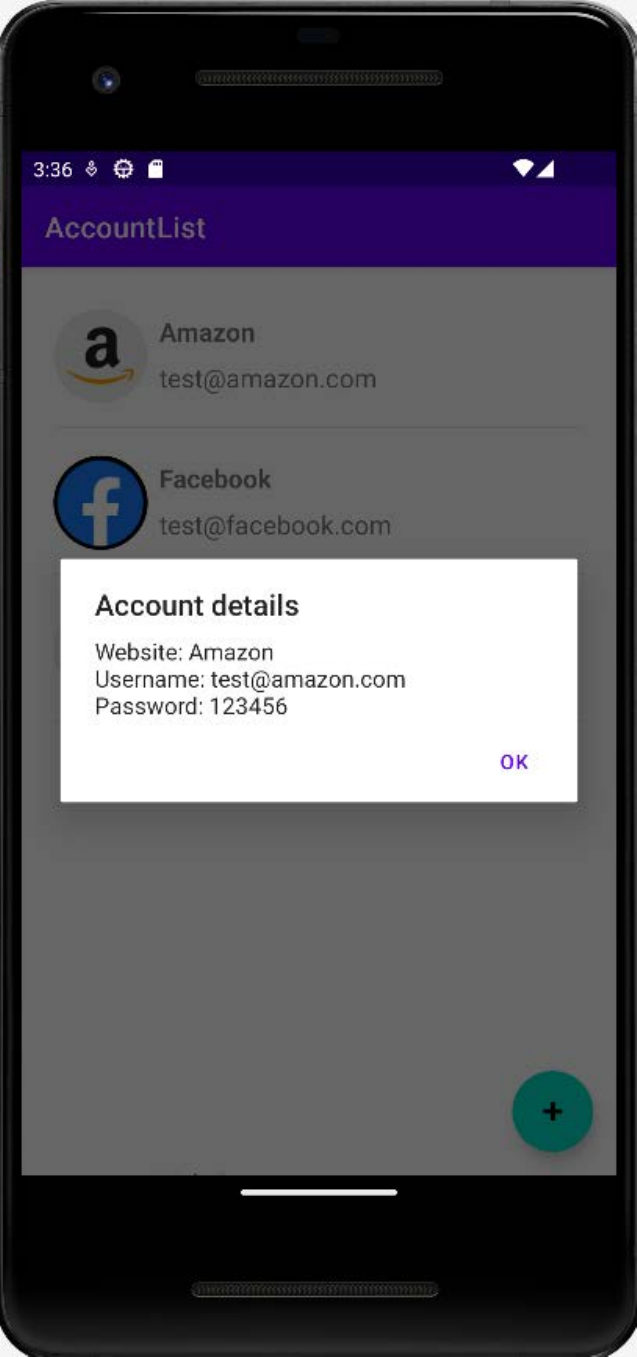}
        \caption{Showing the credentials including the password after successful biometric authentication.}
        \label{fig:decryptedAccount}
    \end{subfigure}

    \caption{Overview of manually viewing credentials in ROSTAM Mobile Application.}
    \label{fig:mainFigure}
\end{figure}

Mobile application of ROSTAM allows the User to view their credentials. This may be needed in a number of situations including for manual login due to security reasons or unsupported platforms (E6-P1). The list of all saved  is shown to a User with URLs and usernames / e-mail addresses (see Figure~\ref{fig:mobileAccountList}). The User can tap on one of them and Mobile App asks to verify User's identity with biometrics (see Figure~\ref{fig:biometricApproval}) before querying the related account’s password from the Server and showing it in clear text after decryption (which is an essential security measure (E6-P2)) as shown in Figure~\ref{fig:decryptedAccount}.

%It acts as a second layer of authentication, providing an additional level of protection against unauthorized access to sensitive information. By requiring the user to confirm their intention to view a password, the application ensures that only authorized individuals have access to the information.
This security feature also helps to prevent accidental exposure of passwords to bystanders (shoulder-surfing attacks) and malicious applications.

%Overall, this feature adds an extra layer of security to the password management process, making it more secure and reliable for the users.

%The following part should be re-written after the final version of ROSTAM is developed.

\subsubsection{Generate Passwords}
\label{sec:generatePassword}

In its current version, ROSTAM only allows manual generation of passwords. This provides users with greater flexibility and control over their passwords. On the other hand, this choice has the risk of leading to weak passwords. We are planning to develop two new features to mitigate this risk: (i) Password Strength Meter, (ii) Random Password Generation.

\subsubsection{Sync credentials}
\label{sec:syncCredentials}
The synchronization of user credentials in ROSTAM happens immediately after the pairing process (explained earlier). Since the credentials are stored on the Server (cloud) in an encrypted form, it is both convenient and secure to access credentials across multiple devices. ROSTAM automatically synchronizes the passwords across devices, ensuring that the User always has access to the most up-to-date data. This eliminates the need for the User to manually update their credentials on each device, saving time and reducing the risk of error. 

Cloud-based synchronization enables users to access their passwords from anywhere with an Internet connection, making it easy to login to their accounts on the go (E8-P1). However, the use of cloud storage for password synchronization requires a high level of security to prevent unauthorized access to user credentials. Password managers using cloud storage should implement effective security measures to protect user data from hackers and cyber threats. As already discussed, ROSTAM addresses these threats by strong encryption with a MasterKey, which is never stored on the Server (cloud).

\subsubsection{Lock Manager}
\label{sec:lockManager}
The Extension allows users to manually or automatically lock the manager. The automatic lock feature is particularly useful for users who may forget to logout from the Extension leaving their credentials open to unauthorized access. The login session expires either after 15 minutes based on recommendations \cite{nistSessionTimeout,owaspSessionTimeout} or when the User closes the browser. This means automatic logout ensuring that sensitive data is protected (E9-P1). 

Additionally, manual lock feature allows users to logout whenever they want, which is particularly useful when using a shared computer or in situations where the User needs to step away from their computer (E9-P2). This prevents unauthorized logins by other even if they have access to the computer. Overall, supporting both manual and automatic lock provides users with greater control and gives them peace of mind that their data is protected.

\subsubsection{Unlock Manager}
\label{sec:unlockManager}
In order to unlock, users are required to login first. Passwordless authentication makes this process convenient and users can quickly and easily unlock without the need to type a username and password. 

%Instead, they can use their biometric data or security key to authenticate themselves, providing a highly secure and convenient way to access their passwords. This removes the risk of password-based attacks, such as phishing, and provides an extra layer of security for users. Overall, unlocking Rostam with FIDO2 without provides users with a highly secure and convenient way to access their credentials, improving the user experience and protecting sensitive information.

\subsection{Recovery of Credentials}
\label{sec:specialcases}
In exceptional cases such as theft or cellphone getting lost/broken, credentials are required to be recovered with a well-designed process since they are not directly available from the cloud.

Assuming that the User now has a new phone, the recovery process can be separated into two parts: (i) Mobile App setup and authentication, (ii) restoring credentials on the new phone. 

Set-up of mobile application on the new phone and User re-registration is similar to the initial setup. A new PubKey and PrivKey gets created on the new phone. But this time, a MasterKey is not generated. Instead a recovery process is initiated to get the MasterKey by the User as described below.

%We assume a user could be still authenticated even without access to his registered phone. We refer readers to recommended measures for FIDO2 passwordless authentication \cite{FIDO2Recovery} and not discuss this procedure in this paper \footnote{In an enterprise setting, offline methods could also be an option.}.  

\begin{figure}
    \includegraphics[width=\columnwidth]{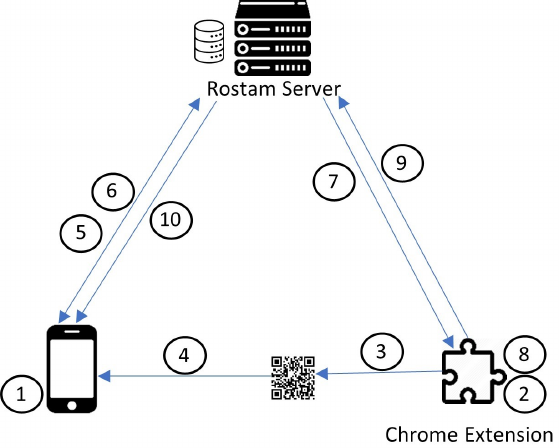}
    \centering
    \small
    \caption{Steps for Recovery in a new phone.} 
    \label{fig:specialcases}
\end{figure}

\begin{figure}
    \includegraphics[width=8cm,keepaspectratio=true]{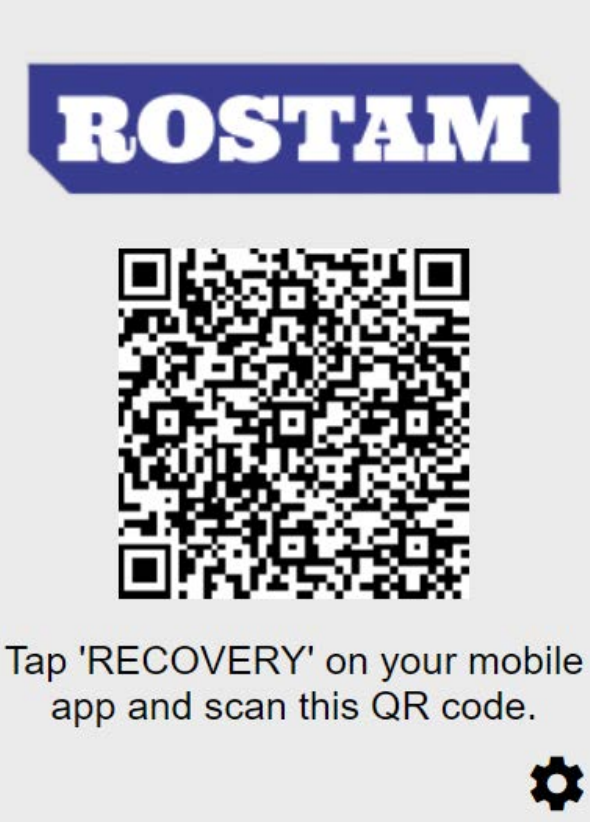}
    \centering
    \small
    \caption{Generated recovery QR Code from token and PubKeyHash in the Extension.}
    \label{fig:QRCodeRecovery}
\end{figure}

Restoring credentials on the new phone is possible as long as the User has access to at least one of the already paired devices. Since the paired device holds the necessary key, which is required to decrypt the encrypted MasterKey stored in ROSTAM's Server, a User can restore credentials in his/her new cellphone as below (Figure~\ref{fig:specialcases}):
\begin{enumerate}
    \item User taps on "RECOVERY" on Mobile App and the camera opens up.
    \item User clicks on "Recovery" button in the Extension.
    \item The Extension generates QR-code of a token and its PubKeyHash concatenated (Figure~\ref{fig:QRCodeRecovery}).
    \item User scans the QR code with the camera.
    \item Mobile App splits the token from PubKeyHash and queries Extension’s PubKey from the Server.
    \item Mobile App concatenates the token with PubKey of the new phone and encrypts it with PubKey of the Extension and saves it to Server in its Recovery field.
    \item The Extension gets the data from Recovery field and also encrypted form of MasterKey from the Server.
    \item The Extension decrypts the data with its PrivKey and extracts the token and mobile device’s PubKey. It compares the token with the local one, and if they match, encrypts MasterKey with the new phone’s PubKey.
    \item The Extension saves the encrypted form of MasterKey in the Server.
    \item The Mobile App gets the encrypted form of MasterKey and after decrypting it with its PrivKey, imports MasterKey to the KeyStore.
\end{enumerate}

\section{Security Analysis}
\label{sec:security}

If the IdP in a Federated Identity System is compromised, this can have significant security implications. But mitigating the impact and restoring security is easier with a passwordless SSO since there is no secret credential that could be stolen and the public keys of users could be restored from a backup. 

On the other hand, in the CM part, if passwords are not stored securely on the Server, then a server compromise has far-reaching and long lasting consequences e.g., all users should be notified to change their passwords as soon as possible. Because of this difference between Federated Identity and CM parts, in the rest of this section, we concentrate on the security of ROSTAM’s CM.

We evaluate the security of ROSTAM’s CM against prominent threats which we divide into two main categories; server-side and client-side. Our focus is on confidentiality and integrity of user data and we leave out availability aspects. We omit a formal security analysis to address it at a later time.

\subsection{Server-side threats}
\label{sec:ssthreats}
Here, there are two possible types of adversaries. One being an insider such as a malicious employee and the other being an outside attacker who breaks into the system. However both scenarios could be addressed at the same time by considering server side as a hostile environment for SPs and minimizing the trust assumptions. We now list some of the well-known attack against the Server and show how strong the design of ROSTAM is against them.
\begin{itemize}
    \item Bypassing authentication: If an attacker successfully bypasses or compromises SSO authentication, to reach user credentials, he needs to pair his device using an already paired user device since the MasterKey is not stored in the Server.

    An attacker capable of bypassing the authentication mechanism but not have a physical access to one of the paired devices may try to exploit the system by deceiving the client-side application, specifically the Extension. The objective is to deceive the Extension into encrypting credentials or the MasterKey using a fake PubKey/MasterKey. Therefore, swapping and transmitting fake keys during the pairing and recovery stages should be avoided.

    One of the primary challenges in our pairing and recovery protocols arises from the unilateral nature of communication with the QR-codes. This methodology does not offer an immediate means of verifying the authenticity of the receiving end. To elaborate, there's no direct mechanism to determine if the encrypted data received from the Server was actually encrypted by the mobile device currently conducting the pairing or recovery process.
    
    To mitigate this vulnerability, we deploy a particular security countermeasure that we consider innovative. For each session, Extension generates a random one-time token. This token is stored exclusively on the client side within the browser. When transmitting the PubKeyHash of the Extension, this one-time token is embedded within the QR Code. Before the encryption, the mobile application must attach this token and then forward the encrypted data to the Server.
    
    In the following interaction, the Extension is now equipped with validating the authenticity of the received encrypted data. It achieves this by verifying that the data originates from the mobile device that previously scanned the QR-code, either during pairing or recovery. This verification is carried out by first decrypting the received data and then comparing the embedded token with the originally generated one. This security measure eliminates the threat for man-in-the-middle attacks, ensuring that only physically present devices can successfully achieve the pairing or recover the data.
    
    \item Access to the Server: The attacker can download the data of all registered users including:
    \begin{enumerate}
        \item Encrypted form of website URLs, usernames and passwords.
        \item PubKeys and encrypted MasterKeys.
    \end{enumerate}
    With this data in hand, an attacker can try a brute force attack. Since MasterKeys are unique for every User, generated randomly without User interaction and strong enough to resist attacks, it is infeasible to crack them. MasterKey provides much better security as compared to the Master Password approach for which incidents has been reported e.g., \cite{LastPassSecurityIncident}. Although policies regarding Master Password choices can improve the security of the keys derived from Master Passwords, there is no guarantee of non-reuse in some other SP.
\end{itemize}
%In case the adversary is an insider, he/she can do the same attacks as an outsider depending on his/her access level. The same conclusions can be true for the consequences of such attack.
\subsection{Client-side threats}
\label{sec:csthreats}
Usually, client-side threats are more challenging to mitigate due to huge number of options for exploiting vulnerabilities on user devices. But they are also less interesting for attackers because the attack does not scale well (compare it with an attack to a server holding thousands or even millions of passwords).

Here, we assume that an attacker has gained access to User’s phone or personal computer either physically or through malware. With this attack PrivKey and/or MasterKey could be targeted but getting these keys might not be easy since they are securely stored. 

In Android version 9.0 or higher \cite{AndroidKeyStore}, KeyStore makes it difficult to extract keys especially in devices which have a security element. For the browser, unfortunately, WebCrypto API specification does not provide useful details on exactly how browsers generate and protect these keys \cite{WebCryptoAPI} but it promises that it would be hard to extract keys. As of writing this article, it is not possible to access an embedded security hardware module directly from Chrome extensions hence we should rely on protection mechanisms provided by WebCrypto API. If a support becomes available, security can be further improved by developing an Extension which can communicate with embedded security hardware e.g., Trusted Platform Module (TPM 2.0) \cite{TPM}. 

Although these protection mechanisms might prevent the attacker to extract the keys, he can still performs decryption with a physical access or by installing a malware which has kernel-level access. For this reason, in the design of ROSTAM, we avoid saving any encrypted data on client’s device. 

The attacker could also attempt to replicate the act of visiting the listed websites in order to get the encrypted passwords from the Server and decrypt them. But, to be able to do so, he needs to break the SSO authentication or use a valid and unexpired session. As stated, the sessions get cleared when browser is closed or after a certain time of User inactivity.

%In conclusion, an attacker in possession of one s (client-side or server-side) will not be able to recover sensitive information easily. It is obvious that an adversary with access to server-side and one of the paired devices can get all data related to specific user.

\section{Evaluation using UDS Framework}
\label{sec:comparison}
In the technical implementation of ROSTAM, we put more work on the CM part. We first looked at current best practices and then brought our own contributions. One of the things we noticed is that many work has tried to prevent server-side attacks by client-side encryption with a secret key generated from a Master Password. In case of a breach on the server-side, the assumption is that Master Password is strong enough and offline brute-force (or dictionary) attacks would fail. One big concern regarding user-chosen strong Master Password is the potential for the User to forget it, which can result in a loss of access to their account and critical data. Although in the past there have been efforts to ease this problem such as memorizing using a technique called spaced repetition \cite{strongmemorablepasswords} or by using graphical passwords \cite{iconbasedpwdgenerator}, it is safe to state that there is still a big burden on users to generate strong master passwords. In general they are far from being as strong as randomly generated keys.

We have seen that in order to increase usability and decrease the possibility of attacks such as keylogging, CMs often cache \cite{TwoCMComparison} Master Passwords temporarily or even permanently. 

An important feature that is missing in some widely used managers is the ability to change or reset Master Password. It is possible that users want to change his/her Master Password in case of compromise.
% reference?

With all these in mind, we come up with our own design which does not have the risks due to predictability of Master Passwords while without losing any benefits already provided by Master Password based methods such as Firefox Sync 2.0. Our encryption scheme uses a key which is same in all devices hence called as MasterKey. For sharing the MasterKey, we use public key cryptography in a novel way we have already discussed.

Using Usability-Deployability-Security (UDS) framework of Joseph Bonneau et al \cite{Replacepasswords}, we compare our proposed method ROSTAM to Firefox Sync 2.0, a typical Master Password based approach. We also make a comparison with TAPAS password manager \cite{tapas} which promises notable improvements over currently available password managers (see Table~\ref{table:encyprion-key-evaluation}):

\subsection{Usability}
\label{sec:usability}
\textbf{Memorywise-Effortless:} ROSTAM, like Tapas, eliminates the need for users to remember a password as long as they have access to one of the paired devices. This is unlike using a Master Password for encryption, which requires users to choose and remember a strong password. 

\textbf{Scalable-for-Users:} Both ROSTAM and Tapas schemes ensure a low cognitive load for users since encryption keys are randomly generated, without burdening the User. 

\textbf{Nothing-to-Carry:} ROSTAM and Tapas fall under the category of Quasi-Nothing-to-Carry since users typically carry their phones with them anyway \footnote{Firefox Synch 2.0 was also previously evaluated as Quasi-Nothing-to-Carry because the authors noted that \textit{''a travelling User will have to carry at least a smartphone: it would be quite insecure to sync one’s passwords with a browser found in a cybercafé \cite{Replacepasswords}".}}.

\textbf{Physically-Effortless:} For both ROSTAM and Tapas, users do not require any effort beyond pressing a button after the one-per-device pairing process (which requires reading a QR-code with a mobile phone).

\textbf{Easy-to-Learn:} ROSTAM, Tapas, and Master Password schemes are all considered easy to learn. Most users are already familiar with QR used in former two.

\textbf{Efficient-to-Use:} ROSTAM is faster to use than Tapas since it does not require users to approve authentication on their phone.

\textbf{Infrequent-Errors:} All three methods minimize errors as users no longer need to type passwords, which can lead to typos.

\textbf{Easy-Recovery-from-Loss:} FireFox Synch 2.0 does not allow easy recovery if master password is forgotten. On the other hand, ROSTAM allows recovery if the User has access to a paired device. This is marked as Quasi-Easy-Recovery-from-Loss due to the recovery steps involved. Tapas does not offer a recovery procedure.

\subsection{Deployability}
\label{sec:deployabality}
\textbf{Accessible:} Similar to Tapas, ROSTAM is marked as Quasi-Accessible since visually impaired users may face difficulties, though modern computers and mobile devices have significantly improved accessibility features.

\textbf{Negligible-Cost-per-User:} None of them needs extra equipment.

\textbf{Server-Compatible:} All three schemes do not necessitate server-side changes.

\textbf{Browser-Compatible:} Both Tapas and ROSTAM require browser extensions, limiting browser compatibility.

\textbf{Mature:} Tapas and ROSTAM are not yet mature, as they only have proof-of-concept implementations.

\textbf{Non-Proprietary:} All three schemes utilize open-source libraries, allowing for public implementation.

\subsection{Security}
\label{sec:Security}
\textbf{Resilient-to-Physical-Observation:} Infrequently-typed Master Password is only partially resilient against shoulder surfing attacks. ROSTAM and Tapas, however, are fully resilient to physical observation.

\textbf{Resilient-to-Targeted-Impersonation:} Master Password can be guessed by an acquaintance, while MasterKey offers greater resilience due to less user interaction and no visible hints for an attacker.

\textbf{Resilient-to-Throttled-Guessing:} Although MasterKey in ROSTAM is strong and difficult to guess, users might still choose weak passwords for SPs.

\textbf{Resilient-to-Unthrottled-Guessing:} Much like the previously mentioned point.

\textbf{Resilient-to-Internal-Observation:} ROSTAM is robust against internal threats, such as keyloggers, providing strong security against internal observation.

\textbf{Resilient-to-Leaks-from-Other-Verifiers:} Master Password is not resilient to leaks from other verifiers, as users may reuse the same key phrase across websites. ROSTAM, on the other hand, is resilient to this type of threat, as users are unaware of MasterKeys.

\textbf{Resilient-to-Phishing:} Master Password is not resilient to phishing, like traditional passwords. However, MasterKey is resilient. Even sophisticated real-time man-in-the-middle attacks are unsuccessful against it.

\textbf{Resilient-to-Theft:} All three schemes provide resilience to theft. In the case of MasterKey, extracting usable data from a stolen device is infeasible.

\textbf{No-Trusted-Third-Party:} Users are not required to trust a third party in any of the schemes, as encryption is performed on the client-side.

\textbf{Unlinkable:} All methods are unlinkable, enhancing user privacy and security.

\begin{table*}[]
\centering
\resizebox{\textwidth}{!}{%
\begin{tabular}{lcccccccccccccccccccccccccccc}
 & & \rotatebox{90}{Memorywise-Effortless}  & \rotatebox{90}{Scalable-for-Users} & \rotatebox{90}{Nothing-to-Carry} & \rotatebox{90}{Physically-Effortless} & \rotatebox{90}{Easy-to-Learn} & \rotatebox{90}{Efficient-to-Use} & \rotatebox{90}{Infrequent-Errors} & \rotatebox{90}{Easy-Recovery-from-Loss} & & \rotatebox{90}{Accessible} & \rotatebox{90}{Negligible-Cost-per-User} & \rotatebox{90}{Server-Compatible} & \rotatebox{90}{Browser-Compatible} & \rotatebox{90}{Mature} & \rotatebox{90}{Non-Proprietary} & & \rotatebox{90}{Resilient-to-Physical-Observation} & \rotatebox{90}{Resilient-to-Targeted-Impersonation} & \rotatebox{90}{Resilient-to-Throttled-Guessing} & \rotatebox{90}{Resilient-to-Unthrottled-Guessing} & \rotatebox{90}{Resilient-to-Internal-Observation} & \rotatebox{90}{Resilient-to-Leaks-from-Other-Verifiers} & \rotatebox{90}{Resilient-to-Phishing} & \rotatebox{90}{Resilient-to-Theft} & \rotatebox{90}{No-Trusted-Third-Party} & \rotatebox{90}{Requiring-Explicit-Consent} & \rotatebox{90}{Unlinkable}\\ \hline
 
 Scheme & \vline &
\multicolumn{8}{c}{\hspace{-0.125in}Usability} & \vline  &
\multicolumn{6}{c}{\hspace{-0.125in}Deployability} & \vline &
\multicolumn{11}{c}{\hspace{-0.125in}Security}
\\ \hline

 TAPAS &  \vline & \full & \full & \prt & \prt & \full & \prt & \full & &  \vline & \prt & \full & \full & & & \full &  \vline & \full & \prt & & & & & \full & \full & \full & \full & \full \\
 Firefox Sync 2.0 (Master Password) &  \vline & \prt & \full & \prt & \prt & \full & \full & \full & &  \vline & \full & \full & \full & & \full & \full &  \vline & \prt & \prt & & & & & \full & \full & \full & \full & \full \\
 ROSTAM (MasterKey) &  \vline & \full & \full & \prt & \prt & \full & \full & \full & \prt &  \vline & \prt & \full & \full & & & \full &  \vline & \full & \full & & & \full & \full & \full & \full & \full & \full & \full \\ \hline

\end{tabular}%
}
\caption{Three schemes evaluated across benefits with respect to Usability, Deployability and Security. Bullets represent provided benefits; empty circles represent partially provided benefits; empty cells means benefits are not provided. }
\label{table:encyprion-key-evaluation}
\end{table*}

\section{Evaluation of ROSTAM's Hybrid Structure}
\label{sec:evaluation}
In this section, we further analyze ROSTAM's hybrid SSO structure using the framework Alaca and Van Oorschot have introduced \cite{SSOAnalysis}. Table~\ref{table:evaluation} also provides properties of OpenID Connect and Firefox Sync 2.0 to facilitate making comparisons.

\textbf{IdP-SP Association Model.}
OpenID Connect uses explicit association (IdP-managed namespace) (A2) and Firefox sync 2.0 or other CMs use Non-attested CM (SP-managed namespace) (A5) Model. As ROSTAM implements both, depending on the SP, it can be either A2 or A5.

\textbf{IdP: User Identity Conveyance Method.} Both conveyance methods are supported by ROSTAM. IdP assertion (G1) is used only for registered RPs and User-to-SP authentication (G3) is used for the rest.

\textbf{SP: User Identity Verification Method.} Hybrid structure of ROSTAM is again relevant here. If the authentication is achieved through OpenID Connect protocol for the RPs, that means IdP Query method (C1) is used. Otherwise, use of CM leads to Local Verification (C2).

\textbf{User to IdP Authentication Type.} Two factor authentication is used to login to ROSTAM SSO (T1b).

\textbf{Multi-Device Usage Model.} ROSTAM adopts Portable hardware token model (M2) as it requires users to carry their phones.

\textbf{B1: Portable-Identity-Across-IdP.} In ROSTAM, users can change their IdP without being required to change or update their SP accounts.

\textbf{B2: No-Device-Setup.} This is partially provided in ROSTAM since Chrome is not installed by default and also the Extension is required to be downloaded and installed on the browser.

\textbf{B3: No-Hardware-Token-Required.} Just like Firefox Sync 2.0 and OpenID Connect, ROSTAM users can authenticate themselves to SPs without any additional hardware token.

\textbf{B4: Resilient-to-Temporary-Service-Outage.}
Unfortunately, User needs to have a persistent connection with the Server of ROSTAM.

\textbf{B5: IdP-Vetting-Not-Needed.} ROSTAM is not required to be registered by any federation operator and therefore our system provides this benefit.

\textbf{B6: SP-Registration-Not-Needed.}
This benefit is provided by ROSTAM since SPs are not required to register.

\textbf{B7: No-SP-Stored-User-Secret.} This benefit is provided only for RPs registered with ROSTAM.

\textbf{B8a: Resilient-to-Client-Leaks.} This benefit means that an attacker should not be able to get any information even by using a keylogger or by recording User keystrokes. In ROSTAM, it is not possible for an attacker to get access to sensitive user data by only accessing a paired device.

\textbf{B8b: Resilient-to-SP-Leaks.} This benefit is provided only for the RPs getting service from the federated identity part, thus ROSTAM partially supports it.

\textbf{B8c: Resilient-to-Third-Party-Leaks (e.g., IdP-leaks).} SSO Passwordless Authentication in ROSTAM ensures that no hashed passwords stored by IdPs are susceptible to exposure.

\textbf{B9: Signals-Assurance-Level-to-SPs.} Currently, it is not provided in ROSTAM.

\textbf{B10: SPs-Can-Filter-IdPs.}
Due to the hybrid nature of ROSTAM, it is partially supported. (SPs for which CM is used cannot distinguish between IdPs.)

\textbf{B11: No-Impersonation-by-Third-Party.} Again, it is partially supported. This time, CM provides this benefit. IdP cannot retrieve plaintext user data since it has no access to MasterKey.

\textbf{B12: Private-Browsing.} Again, partially supported. The IdP has partial knowledge of the SPs that its users authenticate to.

\textbf{B13: Unlinkable-Across-SPs.} For RPs this benefit is not provided. On the other hand, if CM is used, since authentication is performed without involvement of a remote third-party server, ROSTAM offers this benefit.

\textbf{B14: No-Sharing-of-User-Data.} For preregistered SPs it is not provided since ROSTAM is an OpenID Connect based IdP and it provides SPs with access to user profile information. This is not true about CM part and therefore Rostam provides this benefit partially.

As discussed, many of the security benefits are partially provided by ROSTAM. In a sense, this gives more flexibility for users and system administators.

\begin{table*}[]
\centering
\resizebox{\textwidth}{!}{%
\begin{tabular}{lccc}
\cline{2-4}
 & OpenID Connect & Firefox Sync 2.0 & ROSTAM \\ \hline
IdP-SP Association Model & A2 & A5 & A2/A5 \\
IdP: User Identity Conveyance Method & G1 & G3 & G1/G3 \\
SP: User Identity Verification Method & C1 & C2 & C1/C2 \\
User to IdP Authentication   Type & T1 & T2a & T1b \\
Multi-Device Usage Model & M1 & M4 & M2 \\ \hline
B1: Portable-Identity-Across-IdP &  & \full & \full \\
B2: No-Device-Setup & \full & \prt & \prt \\
B3: No-Hardware-Token-Required & \full & \full & \full \\
B4: Resilient-to-Temporary-Service-Outage &  & \full & \\ \hline
B5: IdP-Vetting-Not-Needed & \full & \full & \full \\
B6: SP-Registration-Not-Needed &  & \full & \full \\
B7: No-SP-Stored-User-Secret & \full &  & \prt \\ \hline
B8a: Resilient-to-Client-Leaks &  &  & \full \\
B8b: Resilient-to-SP-Leaks & \full &  & \prt \\
B8c: Resilient-to-Third-Party-Leaks &  &  & \full \\
B9: Signals-Assurance-Level-to-SPs &  &  &  \\
B10: SPs-Can-Filter-IdPs & \full &  & \prt \\
B11: No-Impersonation-by-Third-Party &  & \prt & \prt \\ \hline
B12: Private-Browsing &  & \full & \prt \\
B13: Unlinkable-Across-SPs &  & \full & \prt \\
B14: No-Sharing-of-User-Data &  & \full & \prt \\ \hline

\end{tabular}%
}
\caption{Three schemes compared by design properties and evaluated across Usability, Deployability, Security and Privacy benefits. Bullets represent provided benefits; empty circles represent partially provided benefits; empty cells represent benefits not provided. }
\label{table:evaluation}
\end{table*}

\section{Limitations and future work}
\label{sec:limitations}
In this paper, we have introduced ROSTAM, a novel approach to improve existing web SSO systems. Although ROSTAM offers significant advancements, it comes with its own set of limitations. In this section, we discuss these in detail to provide a comprehensive understanding of the system's constraints and potential areas for future improvements.

\subsection{Application Download Requirement}

A significant limitation is the necessity for users to download and install a mobile application and a browser extension. This requirement may discourage potential users who are reluctant to install additional software on their devices, especially if they are concerned about device performance, storage space, or privacy.

\subsection{Persistent Connection Requirement}

In its current design, ROSTAM demands a continuous connection to the Server. This reliance on connectivity might bring challenges in situations where users experience intermittent or unreliable Internet connections. Furthermore, this dependence on an additional Server might raise concerns about system availability since Server downtime could prevent users from accessing their accounts.

\subsection{Limited Control over Account Creation and Deletion}

For SPs that use the CM of ROSTAM, users must manage passwords by themselves. This limitation could lead to users employing weak or reused passwords, which conflicts with the objective of enhancing security.

\subsection{Vulnerability to Client-Side Malware Attacks}

ROSTAM is unable to protect users from malware targeting client-side device memory. Although login credentials are stored in encrypted form and decrypted only when needed, an attacker or malware listening to the device memory could still potentially access to plaintext credentials. 

\subsection{Dependence on Online Servers for Recovery}

ROSTAM relies on its Server to store and manage account recovery data. If user data is wiped out from the Server whether due to server failure, cyber-attacks, or data corruption, then it is not possible to recover accounts. This reliance on a single point of failure poses a significant risk.

\subsection{Compatibility Issues}

The available implementation may not be compatible with all existing websites, applications, or platforms, limiting widespread adoption. Future development should focus on enhancing compatibility with various systems and platforms to ensure that users can enjoy the benefits of ROSTAM across multiple scenarios.

\subsection{User Acceptance and Trust}

Adoption may be hindered by user reluctance to trust a new and relatively untested authentication system. The success of ROSTAM depends on its ability to gain trust and demonstrate its effectiveness and security compared to traditional authentication methods. This challenge necessitates comprehensive testing, transparent communication, and ongoing user education.

To conclude this section, while ROSTAM presents a promising approach to improve existing methods, it is essential to acknowledge and address its limitations. Further research and development effort should focus on resolving these to enhance security, reliability, and user experience. By addressing these concerns, ROSTAM has the potential to become a powerful and widely adopted authentication solution, providing users with a more secure and usable means of accessing their online accounts.

\section{Conclusion}
We expect a major portion of websites will continue to ask for passwords for a substantial duration of time since for most developers and administrators this is the path of least resistance. While encouraging to adopt passwordless authentication methods to enhance security and user experience is an option, we believe that the real passwordless user experience could be achieved much earlier even if most web sites still use passwords. ROSTAM is a SSO solution we designed and implemented to reflect this position. The notable features of ROSTAM are as follows:

\begin{itemize}
    \item ROSTAM integrated CM and federated identity approaches. It has a dashboard that presents all applications the User can access with a single click after the SSO.
    \item ROSTAM implemented SSO with passwordless authentication. As a result, the User even does not need to memorize a single password.
    \item With ROSTAM, if the Server is hacked, users do not need to change their web site passwords. Most significantly, ROSTAM achieved this critical security aspect without depending on a master password.
    \item In ROSTAM's system model, mobile phone plays a crucial role and acts both as an authenticator as well as the ultimate place all passwords are saved securely. But we also designed a secure recovery method so that all passwords are still reachable in case of a lost, stolen or broken phone. 
\end{itemize}

In the future, we intend on continuing the development of ROSTAM which facilitates passwordless authentication not in the future but at this moment while passwords are still prevalent.

\label{sec:conclusion}

\section{Acknowledgements}
\label{sec:Acknowledgements}
Funding: This research was supported by TUBITAK (The Scientific and Technological Research Council of Turkey) [grant number 3211046].
We would like to express our gratitude to Burak Şahin for his invaluable assistance in testing the system. 



 \bibliographystyle{elsarticle-num} 
 \bibliography{bibfile}

\begin{thebibliography}{10}
\expandafter\ifx\csname url\endcsname\relax
  \def\url#1{\texttt{#1}}\fi
\expandafter\ifx\csname urlprefix\endcsname\relax\def\urlprefix{URL }\fi
\expandafter\ifx\csname href\endcsname\relax
  \def\href#1#2{#2} \def\path#1{#1}\fi

\bibitem{Replacepasswords}
J.~Bonneau, C.~Herley, P.~C.~v. Oorschot, F.~Stajano, \href{https://doi.org/10.1109/SP.2012.44}{The quest to replace passwords: A framework for comparative evaluation of web authentication schemes}, in: Proceedings of the 2012 IEEE Symposium on Security and Privacy, SP '12, IEEE Computer Society, USA, 2012, p. 553–567.
\newblock \href {https://doi.org/10.1109/SP.2012.44} {\path{doi:10.1109/SP.2012.44}}.
\newline\urlprefix\url{https://doi.org/10.1109/SP.2012.44}

\bibitem{SSOAnalysis}
F.~Alaca, P.~C.~V. Oorschot, \href{https://doi.org/10.1145/3409452}{Comparative analysis and framework evaluating web single sign-on systems}, ACM Comput. Surv. 53~(5) (sep 2020).
\newblock \href {https://doi.org/10.1145/3409452} {\path{doi:10.1145/3409452}}.
\newline\urlprefix\url{https://doi.org/10.1145/3409452}

\bibitem{SSOCategories}
A.~Pashalidis, C.~J. Mitchell, A taxonomy of single sign-on systems, in: Proceedings of the 8th Australasian Conference on Information Security and Privacy, ACISP'03, Springer-Verlag, Berlin, Heidelberg, 2003, p. 249–264.

\bibitem{tapas}
D.~McCarney, D.~Barrera, J.~Clark, S.~Chiasson, P.~C. van Oorschot, \href{https://doi.org/10.1145/2420950.2420964}{Tapas: Design, implementation, and usability evaluation of a password manager}, in: Proceedings of the 28th Annual Computer Security Applications Conference, ACSAC '12, Association for Computing Machinery, New York, NY, USA, 2012, p. 89–98.
\newblock \href {https://doi.org/10.1145/2420950.2420964} {\path{doi:10.1145/2420950.2420964}}.
\newline\urlprefix\url{https://doi.org/10.1145/2420950.2420964}

\bibitem{OnlinevsofflineCMUserstudy}
A.~Karole, N.~Saxena, N.~Christin, A comparative usability evaluation of traditional password managers, in: Proceedings of the 13th International Conference on Information Security and Cryptology, ICISC'10, Springer-Verlag, Berlin, Heidelberg, 2010, p. 233–251.

\bibitem{JQuery}
T.~jQuery Foundation, jquery, \url{https://jquery.com/} (2006).

\bibitem{Bootstrap}
I.~Twitter, Bootstrap, \url{https://getbootstrap.com/} (2011).

\bibitem{Flask}
A.~Ronacher, the Flask~community, Flask, \url{https://flask.palletsprojects.com/} (2010).

\bibitem{OpenID}
O.~Foundation, Openid connect, \url{https://openid.net/connect/} (2014).

\bibitem{Authlib}
H.~Yang, Authlib, \url{https://github.com/lepture/authlib} (2016).

\bibitem{OAuth}
T.~O.~W. Group, Oauth 2.0, \url{https://oauth.net/2/} (2012).

\bibitem{Chrome}
G.~LLC, Google chrome, \url{https://www.google.com/chrome/} (2008).

\bibitem{Dexie}
D.~Fahlander, Dexie.js, \url{https://github.com/dexie/dexie} (2013).

\bibitem{MongoDB}
I.~MongoDB, Mongodb, \url{https://www.mongodb.com/} (2007).

\bibitem{Android}
G.~LLC, Android, \url{https://www.android.com/} (2003).

\bibitem{PlayStore}
G.~LLC, Google play store, \url{https://play.google.com/store} (2008).

\bibitem{AndroidKeyStore}
A.~O.~S. Project, \href{https://developer.android.com/training/articles/keystore}{Android keystore system}.
\newline\urlprefix\url{https://developer.android.com/training/articles/keystore}

\bibitem{JavaCrypto}
O.~Corporation, \href{http://bit.ly/3iRNVzD}{Java Cryptography Extension (JCE)}.
\newline\urlprefix\url{http://bit.ly/3iRNVzD}

\bibitem{SHA}
Q.~Dang, Secure hash standard (2015-08-04 2015).
\newblock \href {https://doi.org/https://doi.org/10.6028/NIST.FIPS.180-4} {\path{doi:https://doi.org/10.6028/NIST.FIPS.180-4}}.

\bibitem{ChromeWebStore}
G.~LLC, \href{https://chrome.google.com/webstore}{Chrome web store}.
\newline\urlprefix\url{https://chrome.google.com/webstore}

\bibitem{PMUseCases}
J.~Simmons, O.~Diallo, S.~Oesch, S.~Ruoti, \href{https://doi.org/10.1145/3485832.3485889}{Systematization of password manageruse cases and design paradigms}, in: Annual Computer Security Applications Conference, ACSAC '21, Association for Computing Machinery, New York, NY, USA, 2021, p. 528–540.
\newblock \href {https://doi.org/10.1145/3485832.3485889} {\path{doi:10.1145/3485832.3485889}}.
\newline\urlprefix\url{https://doi.org/10.1145/3485832.3485889}

\bibitem{nistSessionTimeout}
{National Institute of Standards and Technology}, \href{https://pages.nist.gov/800-63-3/sp800-63b.html#aal3reauth}{Digital identity guidelines: Authentication and lifecycle management}, NIST Special Publication 800-63B.
\newline\urlprefix\url{https://pages.nist.gov/800-63-3/sp800-63b.html#aal3reauth}

\bibitem{owaspSessionTimeout}
OWASP, \href{https://cheatsheetseries.owasp.org/cheatsheets/Session_Management_Cheat_Sheet.html#session-expiration}{Session management cheat sheet}, OWASP Cheat Sheet Series.
\newline\urlprefix\url{https://cheatsheetseries.owasp.org/cheatsheets/Session_Management_Cheat_Sheet.html#session-expiration}

\bibitem{LastPassSecurityIncident}
LastPass, \href{http://bit.ly/3Db61U6}{Lastpass august 25, 2022 security incident}, [Online; accessed 18-January-2023] (2022).
\newline\urlprefix\url{http://bit.ly/3Db61U6}

\bibitem{WebCryptoAPI}
W.~W.~W. Consortium, \href{https://www.w3.org/TR/WebCryptoAPI/}{Web cryptography api}, [Online; accessed 18-January-2023] (2014).
\newline\urlprefix\url{https://www.w3.org/TR/WebCryptoAPI/}

\bibitem{TPM}
T.~C. Group, \href{http://bit.ly/3D81KR0}{Trusted platform module (tpm 2.0) specification}, Trusted Computing Group website, [Accessed 18-January-2023] (2015).
\newline\urlprefix\url{http://bit.ly/3D81KR0}

\bibitem{strongmemorablepasswords}
J.~Bonneau, S.~Schechter, Towards reliable storage of 56-bit secrets in human memory, in: Proceedings of the 23rd USENIX Conference on Security Symposium, SEC'14, USENIX Association, USA, 2014, p. 607–623.

\bibitem{iconbasedpwdgenerator}
R.~Biddle, M.~Mannan, P.~C. van Oorschot, T.~Whalen, \href{https://doi.org/10.1109/TIFS.2011.2116781}{User study, analysis, and usable security of passwords based on digital objects}, Trans. Info. For. Sec. 6~(3) (2011) 970–979.
\newblock \href {https://doi.org/10.1109/TIFS.2011.2116781} {\path{doi:10.1109/TIFS.2011.2116781}}.
\newline\urlprefix\url{https://doi.org/10.1109/TIFS.2011.2116781}

\bibitem{TwoCMComparison}
R.~Zhao, C.~Yue, K.~Sun, Vulnerability and risk analysis of two commercial browser and cloud based password managers 1 (2013) 1--15.

\end{thebibliography}





% \end{thebibliography}
\end{document}